\documentclass[twocolumn,showpacs,preprintnumbers,amsmath,amssymb,unsortedaddress,
]{revtex4-1}
\usepackage{graphicx}
\usepackage{dcolumn}
\usepackage{bm}
\usepackage{soul}
\usepackage{color}
\usepackage{epstopdf}
\usepackage[version=3]{mhchem}
\usepackage{lipsum}
\usepackage[outercaption]{sidecap}
\usepackage{floatrow}
\usepackage{amsmath}
\usepackage{lipsum}
\begin{document}

\preprint{}

\title{Theory of Activated Glassy Dynamics in Randomly Pinned Fluids}
\thanks{Corresponding author:}%

\author{Anh D. Phan$^{1,4,5}$ and Kenneth S. Schweizer$^{2,3,4}$}
\email{kschweiz@illinois.edu}

\affiliation{Department of Physics$^1$, Materials Science$^2$, Chemistry$^3$, Frederick Seitz Materials Research Laboratory$^4$, University of Illinois, Urbana, Illinois 61801, USA} 

\address{Institute of Physics$^5$, Vietnam Academy of Science and Technology, 10 Dao Tan, Hanoi, Vietnam}


\date{\today}

\begin{abstract}
We generalize the force-level, microscopic, Nonlinear Langevin Equation (NLE) theory and its elastically collective generalization (ECNLE theory) of activated dynamics in bulk spherical particle liquids to address the influence of random particle pinning on structural relaxation. The simplest neutral confinement model is analyzed for hard spheres where there is no change of the equilibrium pair structure upon particle pinning.  As the pinned fraction grows, cage scale dynamical constraints are intensified in a manner that increases with density. This results in the mobile particles becoming more transiently localized, with increases of the jump distance, cage scale barrier and NLE theory mean hopping time; subtle changes of the dynamic shear modulus are predicted. The results are contrasted with recent simulations. Similarities in relaxation behavior are identified in the dynamic precursor regime, including a roughly exponential, or weakly supra-exponential, growth of the alpha time with pinning fraction and a reduction of dynamic fragility. However, the increase of the alpha time with pinning predicted by the local NLE theory is too small, and severely so at very high volume fractions. The strong deviations are argued to be due to the longer range collective elasticity aspect of the problem which is expected to be modified by random pinning in a complex manner.  A qualitative physical scenario is offered for how the three distinct aspects that quantify the elastic barrier may change with pinning. ECNLE theory calculations of the alpha time are then presented based on the simplest effective-medium-like treatment for how random pinning modifies the elastic barrier. The results appear to be consistent with most, but not all, trends seen in recent simulations. Key open problems are discussed with regards to both theory and simulation.  
\end{abstract}

\maketitle


\section{Introduction}
Understanding the physical mechanisms underlying the glass transition remains a grand challenge \cite{1,2,3}. When liquids are cooled their structural relaxation time dramatically increases by 14 or more decades before the system falls out of equilibrium heralding kinetic vitrification. Simulations typically probe 4-6 decades of the initial slowing down--the so-called dynamical precursor regime. Many theories have been advanced based on qualitatively distinct hypotheses \cite{1,2,3,4,5,6,7,8,9,10,11,12,13,14,15,16,17,18,19}. These include approaches that relate glassy dynamics to equilibrium thermodynamics such as the entropy crisis Adams-Gibbs model \cite{3,4} and Random First Order Transition (RFOT) theory \cite{13,14}, and explicitly dynamical approaches such as mode coupling theory \cite{18}, dynamic facilitation \cite{15,16}, correlated strings \cite{19}, and local cage scale hopping \cite{9,10} coupled with longer range collective elasticity \cite{7,8,11}. 

In an effort to critically test theoretical ideas, a recent theme has been to employ simulation to probe the sensitivity of glassy dynamics to boundary conditions \cite{3,20}.  A bulk realization of this idea introduces internal constraints, the so-called random pinning protocol \cite{21}. Here, a subset of particles are randomly fixed in space in a manner that does not change the structural pair correlations, so-called neutral confinement \cite{21,22,23,24,25,26,27,28}. Such random pinning leads to slower relaxation in a manner that depends strongly on the pinning fraction and system temperature or density. Although many interesting simulation results for different idealized spherical particle models have been obtained \cite{22,23,24,25,26,27,28}, it seems fair to say this body of work has not provided a definitive test of competing theories for at least two reasons. (i) Simulations only probe the dynamical precursor regime where there are non-universal crossover effects that often are not well understood. (ii) Most theories do not make testable quantitative predictions for how random pinning changes activated dynamics, a limitation that must be addressed to make definitive progress \cite{25}. Beyond the basic physics motivation, randomly pinned systems are toy models of real quenched porous media, including colloidal suspensions with particles pinned using optical tweezers \cite{29}. 

The present work is motivated by both basic physics and porous media considerations. We aim to construct a theory for the effect of random pinning by extending the elastically collective nonlinear Langevin equation (ECNLE) approach \cite{7,8,30} of activated dynamics in 1-component liquids. ECNLE theory is formulated at the level particles and forces, and relates structure and thermodynamics to relaxation. It has successfully predicted, often with no adjustable parameters, relaxation in colloidal \cite{8,30}, molecular \cite{7,8} and polymeric \cite{31,32} systems. General and material-specific aspects of relaxation over 14-16 decades in time have been analyzed.

Figure 1 sketches the physical ideas of ECNLE theory. Building on a simple ("{na\"ive}") version of ideal mode coupling theory (NMCT \cite{13,33}), a stochastic trajectory level approach for cage scale single particle barrier hopping was constructed, NLE theory \cite{10,33}. Though successful for the initial few decades of slow dynamics in the precursor regime \cite{20,24}, NLE theory breaks down at lower temperatures and higher densities where it under predicts (eventually severely) the relaxation time \cite{7,8}. The physical reason has been argued to be associated with the need to create a small amount of local free volume via cage dilation to allow large amplitude hopping events to occur \cite{7,11,30}. This cage dilation can be realized via a spontaneous collective elastic fluctuation of particles outside the cage which is quantified via a radially-symmetric displacement field with a characteristic amplitude and spatial form. The alpha relaxation event then becomes of mixed local-nonlocal character whereby the longer range elastic fluctuation contributes to the activation barrier and serves as a facilitating process to allow irreversible local re-arrangement. Above (below) a characteristic liquid packing fraction (temperature), the collective elastic component dominates the growth of the relaxation time \cite{7,8}. This crossover is predicted to occur close to the empirically deduced (via extrapolation) "mode coupling transition" (MCT) volume fraction ($\sim$0.58) or temperature ($\sim$1.1-1.3$T_g$, where $T_g$ is the experimental vitrification temperature) where activated dynamics is already important and not contained in ideal MCT. We emphasize that this empirical MCT crossover is not the ab initio computed ideal MCT transition which occurs at a significantly higher (lower) temperature (packing fraction) and is more properly thought of as an “onset” condition \cite{3,7}.

This present article presents our initial attempt to generalize ECNLE theory to pinned-mobile systems. We consider a fluid of identical hard spheres with a fraction   randomly pinned. Pinning intensifies confining forces on the cage scale as described via the "dynamic free energy" of NLE theory, and also introduces changes of the emergent shear rigidity and nature of the facilitating displacement field fluctuations required to allow a large amplitude hopping event to be realized. Physically, we expect pinning has strong consequences on all mobile particle dynamical properties as encoded in the dynamic free energy. How it modifies the collective elastic effects are analyzed in an effective medium framework. 

Section II briefly reviews NLE and ECNLE theories for bulk homogeneous 1-component and binary mixture sphere fluids. The NLE approach is extended to treat the effect of pinning in section III. Numerical calculations of the dynamic localization length, jump distance, shear modulus, entropic barrier, and mean alpha relaxation time are presented. The results are compared to recent simulation studies, and agreements and disagreements are identified. An approximate analytic analysis is performed and the derived results provide physical insight to the numerical results. Predictions for the alpha relaxation time of an effective medium extension of ECNLE theory are presented in section IV. The paper concludes in section V with a brief summary and discussion. Three Appendices provide technical details of the theoretical development, implementation and analytic analysis. 
\section{Dynamical Theories of Bulk Liquids}
As relevant background, we recall NLE and ECNLE theories of 1 and 2 component fluids in the absence of pinning \cite{7,34}.  All applications below are for hard spheres and the required structural correlations are computed with Percus-Yevick (PY) integral equation theory \cite{35}.
\subsection{Single-Component Fluid: NLE Theory}
We consider a hard sphere (diameter $d$) fluid of volume fraction $\Phi=\pi\rho d^3/6$. Adopting a {na\"ive} mode coupling approach based on density fluctuations as the slow variable, the force-force time correlation function experienced by a tagged particle in Fourier space is \cite{13,33,36,37}

\begin{eqnarray}
\left\langle \mathbf{f}(0).\mathbf{f}(t) \right\rangle = (k_BT)^2\rho\int\frac{d\mathbf{q}}{(2\pi)^3}\left|\mathbf{M}(q)\right|^2S(q)\Gamma_s(q,t)\Gamma_c(q,t),\nonumber\\
\end{eqnarray}
where $k_B$ is the Boltzmann  constant, $T$ is temperature, $\beta = 1/(k_BT)$, $S(q)$ is the collective static structure factor, $q$ is the wavevector, the effective force is $\mathbf{M}(q)=\mathbf{q}C(q)$,  $C(q)=\rho^{-1}\left[1-S^{-1}(q) \right]$ is the direct correlation function, and $\Gamma_s(q,t)= \left< e^{i\mathbf{q}\left(\mathbf{r}(t)-\mathbf{r}(0) \right)} \right>$ and $\Gamma_c(q,t)=S(q,t)/S(q)$ are the normalized (at $t=0$) single and collective dynamic propagators, respectively. The kinetically arrested state is treated as an Einstein glass corresponding to particles isotropically localized on a length scale $r_L$. In the long time limit, the dynamic propagators become Debye-Waller factors \cite{10,33}:
\begin{eqnarray}
\Gamma_s(q,t\rightarrow \infty) &=& e^{-q^2r_L^2/6}, \nonumber\\
\Gamma_c(q,t\rightarrow \infty) &=& e^{-q^2r_L^2/6S(q)}.
\end{eqnarray}

The collective contribution includes the deGennes narrowing effect. Its form is motivated by from the short time collective density fluctuation propagator, $\Gamma_c(q,t) = e^{-q^2D_st/S(q)}$ where $D_s = k_BT/\zeta_s$ is the short time self diffusion constant. Single particle localization in the long time limit is enforced via the replacement $6D_st = \left\langle (\mathbf{r}(t\rightarrow \infty)-\mathbf{r}(0))^2\right\rangle \rightarrow r_{L}^2$. The self-consistent expression for $r_L$ follows from the "spring constant" $K_\infty = \beta\left\langle \mathbf{f}(0).\mathbf{f}(t) \right\rangle$ which obeys $K_\infty(r_L)r_L^2=3k_BT$, thereby yielding the NMCT self-consistent localization equation \cite{33}:
\begin{eqnarray}
\frac{1}{r_L^2} = \frac{\rho}{9}\int d\mathbf{q} \frac{\left|\mathbf{M}(q)\right|^2S(q)}{\left(2\pi \right)^3}\exp\left(-\frac{q^2r_L^2}{6}\left[1+S^{-1}(q) \right] \right).\nonumber\\
\label{eq:local}
\end{eqnarray}

An idealized localized state is predicted at $\Phi_c \approx 0.43$ \cite{33}.

\begin{figure}[htp]
\center
\includegraphics[width=8cm]{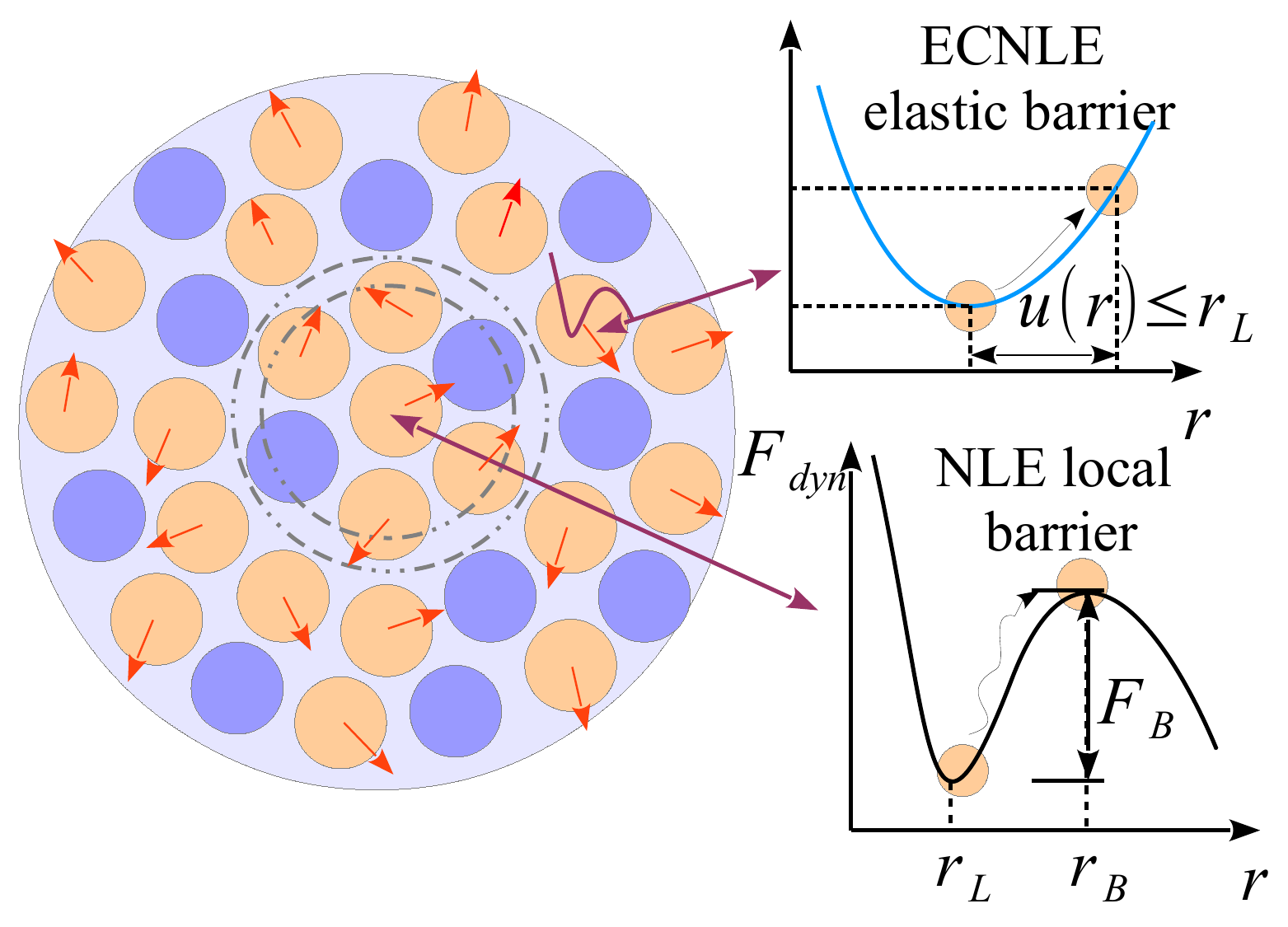}
\caption{\label{fig:1}(Color online) Schematic illustration of the ideas of NLE and ECNLE theory for the pinned-mobile particle system. Violet and orange spheres correspond to pinned and mobile particles, respectively.}
\end{figure}

NLE theory goes beyond ideal NMCT to predict activated single particle stochastic trajectories described at the level of an angularly-averaged scalar dynamic displacement, $r(t)$ , of a tagged particle from its initial position. In the overdamped limit one has \cite{10,33}:

\begin{eqnarray}
\zeta_s\frac{dr(t)}{dt}-\frac{\partial F_{dyn}(r(t))}{\partial r(t)}+ \xi(t) = 0,
\label{eq:4}
\end{eqnarray}
where $\xi(t)$ is the white noise random force corresponding to the short time Fickian diffusion process. $F_{dyn}(r)$ is the "dynamic free energy", the gradient of which describes an effective force on a tagged particle due to the surrounding particles. It is given by \cite{10,33,37}:
\begin{widetext}
\begin{eqnarray}
\frac{F_{dyn}(r)}{k_BT} &=& -3\ln\frac{r}{d}-\rho\int\frac{d\mathbf{q}}{(2\pi)^3}\frac{\left|\mathbf{M}(q)\right|^2S(q)}{q^2\left[1+S^{-1}(q) \right]}\exp\left[-\frac{q^2r^2}{6}\left(1+S^{-1}(q) \right) \right] \equiv \frac{F_{ideal}(r)}{k_BT}+\frac{F_{cage}(r)}{k_BT}.
\label{eq:free}
\end{eqnarray}
\end{widetext}
The leading term favors the fluid state and the second term corresponds to a trapping potential due to interparticle forces which favors localization. If the noise term in Eq.(\ref{eq:4}) is dropped, the NMCT ideal glass transition is recovered. For $\Phi > \Phi_c$, $F_{dyn}(r)$ has a minimum at $r_L$ (which obeys Eq.(\ref{eq:local})) and a barrier at displacement of $r_B$ of height  $F_B$, as sketched in Fig.\ref{fig:1}.

\subsection{Single-Component Fluid: ECNLE Theory}
ECNLE theory introduces facilitating longer-range collective elastic fluctuations which are argued to be essential for allowing cage scale hopping at sufficiently low temperature or high density \cite{7,11,30}. The elastic fluctuation is described by a strain or displacement field outside the cage radius (defined from the first minimum of the pair correlation function, $g(r)$) of:
\begin{eqnarray}
u(r) = \Delta r_{eff}\left(\frac{r_{cage}}{r}\right)^2, \qquad r \geq r_{cage}
\label{eq:elastic1}
\end{eqnarray}
where $r_{cage}\approx 1.3-1.5d$, and the cage dilation amplitude is of order or smaller than the transient localization length and is given by \cite{7,30}:
\begin{eqnarray}
\Delta r_{eff} &=& \frac{3}{r_{cage}^3}\left[\frac{r_{cage}^2\Delta r^2}{32}-\frac{r_{cage}\Delta r^3}{192}+\frac{\Delta r^4}{3072} \right] \nonumber\\ 
&\approx& \frac{3}{32}\frac{\Delta r^2}{r_{cage}},
\label{eq:effective}
\end{eqnarray}
Here, $\Delta r = r_B - r_L$ is the microscopic jump distance (Fig.\ref{fig:1}). The elastic energy cost is then:
\begin{eqnarray}
F_e &=& 2\pi\int_{r_{cage}}^\infty dr r^2\rho g(r) K_0 u^2(r) \nonumber\\
 &=& 12\Phi K_0\Delta r_{eff}^2\left(\frac{r_{cage}}{d}\right)^3,
\label{eq:elastic2}
\end{eqnarray}
where $K_0 = 3k_BT/r_L^2$ is the curvature of $F_{dyn}(r)$ at $r = r_L$. The final result of Eq.(\ref{eq:elastic2}) assumes $g(r) = 1$ outside the cage, which is a benign simplification for hard spheres \cite{7}. 

The alpha process is viewed as a mixed local-nonlocal activated event with a total barrier composed of cage (NLE theory) and collective elastic contributions, $F_{total} = F_B + F_e$. For $\Phi \leq 0.54$, the latter is small or negligible compared to the local barrier. One measure of a dynamic crossover is when the rate of increase of the elastic and local barriers with increasing volume fraction (slope) are equal; this criterion yields \cite{7} a crossover at $\Phi_x\sim 0.575$. Another measure of crossover, popular in the analysis of experiments and simulations [3,38,39], is to empirically fit alpha time data to a mode coupling critical power law expression. Implementing this procedure for ECNLE numerical calculations yields \cite{7} an "empirical MCT crossover" at $\Phi_c\sim 0.58-0.59$. The latter corresponds to $\sim 5-6$ decades of growth of the alpha time in the dynamic precursor regime, the maximum range typically probed in simulation. The kinetic glass transition, corresponding to a $\sim 14$ decade growth of the alpha time with volume fraction in hard spheres (or an alpha time of 100 s for thermal liquids), is predicted \cite{8} to occur at $\Phi_g \approx 0.61$, where $F_e$ is modestly larger than $F_B$.

For bulk liquids one can qualitatively compare trends of hard sphere systems with those of thermal supercooled liquids of spherical particles by identifying volume fraction with inverse temperature \cite{40}. This connection is also in the spirit of a quantitative mapping from an effective hard sphere fluid to a molecular or polymer liquid in the ECNLE framework \cite{8}. Whether such a connection for pinned-mobile fluids is reliable for all aspects of how pinning slows dynamics is not obvious. We return to this below.

\subsection{Two-Component Liquids}
NMCT and NLE theories for the pinned-mobile particle system are constructed via taking a special limit of the general 2-component fluid mixture NMCT and NLE theories discussed previously \cite{34}. Key technical details are collected in Appendices A and B. 

Binary mixture NMCT predicts ideal kinetic glass arrest via the individual, species-dependent long time mean-square displacements    $\left\langle (r_i(t\rightarrow \infty)-r_i(0))^2\right\rangle = r_{L,i}^2$, where $i$ denotes the species $i$ ($i=1,2$). The latter obeys coupled self-consistent equations \cite{34}:
\begin{eqnarray}
\frac{3k_BT}{2} = \left\langle \mathbf{f}_{i}(t\rightarrow \infty)\mathbf{f}_{i}(0) \right\rangle \frac{r_{L,i}^2}{2},
\label{eq:NMCT}
\end{eqnarray}
where $\mathbf{f}_{i}(t)$ is total effective force acting on the species $i$ at time $t$. The required force-force time correlations are \cite{34}:
\begin{widetext}
\begin{eqnarray}
\left\langle \mathbf{f}_{i}(t)\mathbf{f}_{i}(0) \right\rangle = \frac{k_B^2T^2}{3}\int\frac{d\mathbf{q}}{(2\pi)^3}q^2\sum_{j,k=1}^{2}c_{ij}(q)\sqrt{\rho_j\rho_k}S_{jk}(q,t)c_{ki}(q)\Gamma_{s,i}(q,t),
\label{eq:force-force}
\end{eqnarray}
\end{widetext}
where $\rho_i$ is the site number density of species $i$, $S_{ij}(q)$ and $c_{ij}(q)$ are the dimensionless collective structure factor and direct correlation function between species $i$ and $j$, respectively. In long time limit, the Debye-Waller factor $\Gamma_{s,i}(q,t\rightarrow \infty) = e^{-q^2r_{L,i}^2/6}$ describes a localized single particle. Its collective analog is more complicated for a binary mixture. The derivation is based on a short time analysis of $S_{ij}(q,t)$ which obeys \cite{34,41,42}
\begin{eqnarray}
\frac{d}{dt}\mathbf{S}(q,t) = -q^2\mathbf{H}(q)\mathbf{S}^{-1}(q)\mathbf{S}(q,t),
\label{eq:hydro}
\end{eqnarray}
where $H_{ij}(q) = (k_BT/\zeta_{s,j})\delta_{ij}$, $\zeta_{s,j} = k_BT/D_{s,j}$ is the short time friction constant for the component $j$, $D_{s,j}$ is the short time self diffusion coefficient, and in matrix form $\mathbf{S}^{-1}(q) = (\mathbf{I} - \mathbf{C}^*)^{-1}$ with $C^*_{ij} = \rho_ic_{ij}(q)$. Equation (\ref{eq:hydro}) then becomes:
\begin{eqnarray}
\frac{d}{dt}\mathbf{S}(q,t) &=& -{\Omega}(q)\mathbf{S}(q,t), \nonumber\\
\Omega_{ij}(q) &=& \frac{k_BT}{\zeta_{s,i}}q^2\left(\delta_{ij} - \rho_ic_{ij}(q) \right).
\label{eq:matrix}
\end{eqnarray}

Straightforward calculation (see \cite{34}, and Appendix A) yields analytic expressions for $S_{ij}(q,t)$. The collective-Debye Waller factors then follow via the binary mixture analog of the 1-component system long time replacement relation \cite{34}: $6k_BTt/\zeta_{s,j}\rightarrow r_{L,j}^2$ and $\zeta_{s,j}/\zeta_{s,i} = r_{L,i}^2/r_{L,j}^2$, which closes the theory for $r_{L,1}$ and $r_{L,2}$. The latter, along with a standard factorization approximation, yields the dynamic elastic shear modulus \cite{34,41}

\begin{eqnarray}
G &=& \frac{k_BT}{60\pi^2}\sum_{\gamma_1,\gamma_2=1}^{2}\sum_{\gamma_3,\gamma_4=1}^{2}\int_{0}^{\infty}dq q^4\frac{dc_{\gamma_1\gamma_3}(q)}{dq}\frac{dc_{\gamma_2\gamma_4}(q)}{dq}\nonumber\\
&\times&
S_{\gamma_1\gamma_2}(q,t\rightarrow\infty)S_{\gamma_3\gamma_4}(q,t\rightarrow\infty).
\label{eq:shear}
\end{eqnarray}

For a binary liquid, a 2-dimensional dynamic free energy surface can be constructed [43]. However, this is not necessary for the pinned-mobile system since only one species moves. Thus, as discussed below, one can go directly from the NMCT level binary mixture description to the analogous NLE theory in a manner identical to how this is executed for a 1-component system \cite{10,33}. Having done that, the extension of ECNLE theory to the pinned-mobile system can be performed within the well-established 1-component dynamical framework \cite{7}.
\section{NMCT and NLE Theories of the Pinned-Mobile System}
\subsection{Formulation}
The pinned-mobile system under neutral confinement obeys $c_{11}(r)=c_{12}(r)=c_{22}(r)=c(r)$, where the subscript 1 and 2 indicate mobile and pinned particles, respectively, and $c(r)$ is the 1-component hard sphere fluid analog. The density of mobile and pinned particles are $\rho_1=\rho(1-\alpha)$ and $\rho_2=\rho\alpha$. Because pinned particles are immobile, their localization length is zero. This constraint is implemented in the 2-component mixture NMCT by letting $\zeta_{s,2} \rightarrow \infty$, which implies $\Omega_{22}(q) = 0$ and $\Omega_{21}(q) = 0$ in Eq.(\ref{eq:matrix}) (see Appendix A). One can then derive (see Appendix B) a single NMCT localization relation for the mobile species as:

\begin{eqnarray}
\frac{9}{r^2_{L1}}&=&\int \frac{d\mathbf{q}}{(2\pi)^3}q^2e^{-q^2r_{L1}^2/6}\left[\frac{c(q)S_{12}}{\rho_1\left(1-\rho_1c(q) \right)}\right. \nonumber\\
&+&\left.  \frac{\rho_1 c(q)^2}{1-\rho_1c(q)}e^{-q^2r_{L1}^2(1-\rho_1c(q))/6} \right].
\label{eq:rL}
\end{eqnarray}

The corresponding NLE description and dynamic free energy for the mobile species is constructed from Eq.(\ref{eq:rL}) exactly as done for a 1-component system. The result is
\begin{eqnarray}
F_{dyn}(r_1) &=& -3\ln r_1 - \int \frac{d\mathbf{q}}{(2\pi)^3}\left[\frac{c(q)S_{12}(q)e^{-q^2r_1^2/6}}{\rho(1-\alpha)\left[1-\rho(1-\alpha) c(q) \right]} \right.\nonumber\\
& +&\left.\frac{\rho(1-\alpha)c^2(q)e^{-q^2r_1^2\left[2-\rho(1-\alpha)c(q)\right]/6}}{\left[1-\rho(1-\alpha)c(q)\right] \left[2- \rho(1-\alpha)C(q)\right]} \right].
\label{eq:1}
\end{eqnarray}

The first term in the square brackets of Eqs.(\ref{eq:rL}) and (\ref{eq:1}) arises from forces between pinned and mobile particles. There is no Debye-Waller-like factor for the former since $r_{L2}=0$, and this term vanishes if the pinned particle fraction is zero since $S_{12}(q)\rightarrow 0$. The second term arises from forces between pairs of mobile particles. Setting the derivative of the dynamic free energy to zero yields Eq.(\ref{eq:rL}), by construction \cite{34}.
\subsection{Numerical Results: Lengths Scales, Barrier and Shear Modulus}
Before implementing the theory, we note that beyond a critical pinning fraction one expects the “accessible free volume” for the mobile particle motion becomes “de-percolated”. Since NLE theory is local and barriers are finite below random close packing, it does not capture this larger length scale effect. Thus, how high in $\alpha$ NLE theory is reliable is unknown. Simulations of various sphere models \cite{23,24,25,26,27,28} typically explore pinning fractions up to $\alpha\sim 0.1-0.2$, although simulations of pinned-mobile water models \cite{44} extend to $\alpha\sim 0.5$ and still find relaxation and diffusion. We perform NLE theory calculations that fall in between these limits.

\begin{figure}[htp]
\center
\includegraphics[width=8cm]{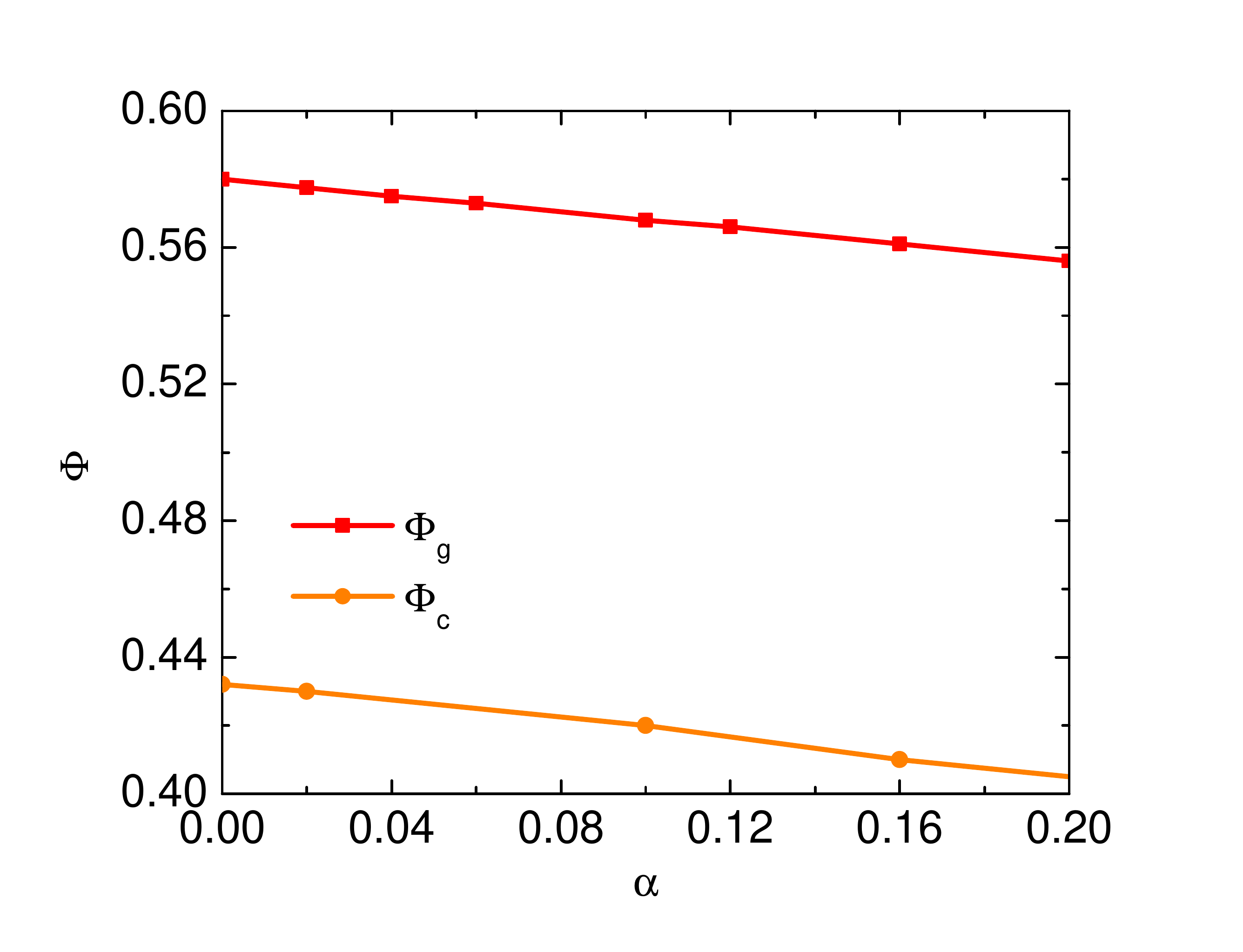}
\caption{\label{fig:2}(Color online) The ideal NMCT ($\Phi_g$) and dynamical arrest ($\Phi_c$) volume fraction versus pinning fraction. The latter volume fraction is defined as when the mean alpha time of the pinned-mobile system equals its pure bulk counterpart at $\Phi=0.58$.}
\end{figure}

Figure \ref{fig:2} shows NMCT calculations of how the ideal glass transition, which is the initial dynamic crossover to the emergence of a barrier in NLE theory, changes with pinning fraction. The onset volume fraction decreases, roughly linearly, by $\sim$ 10$\%$ as $\alpha$ grows to 20$\%$.

\begin{figure}[htp]
\center
\includegraphics[width=8cm]{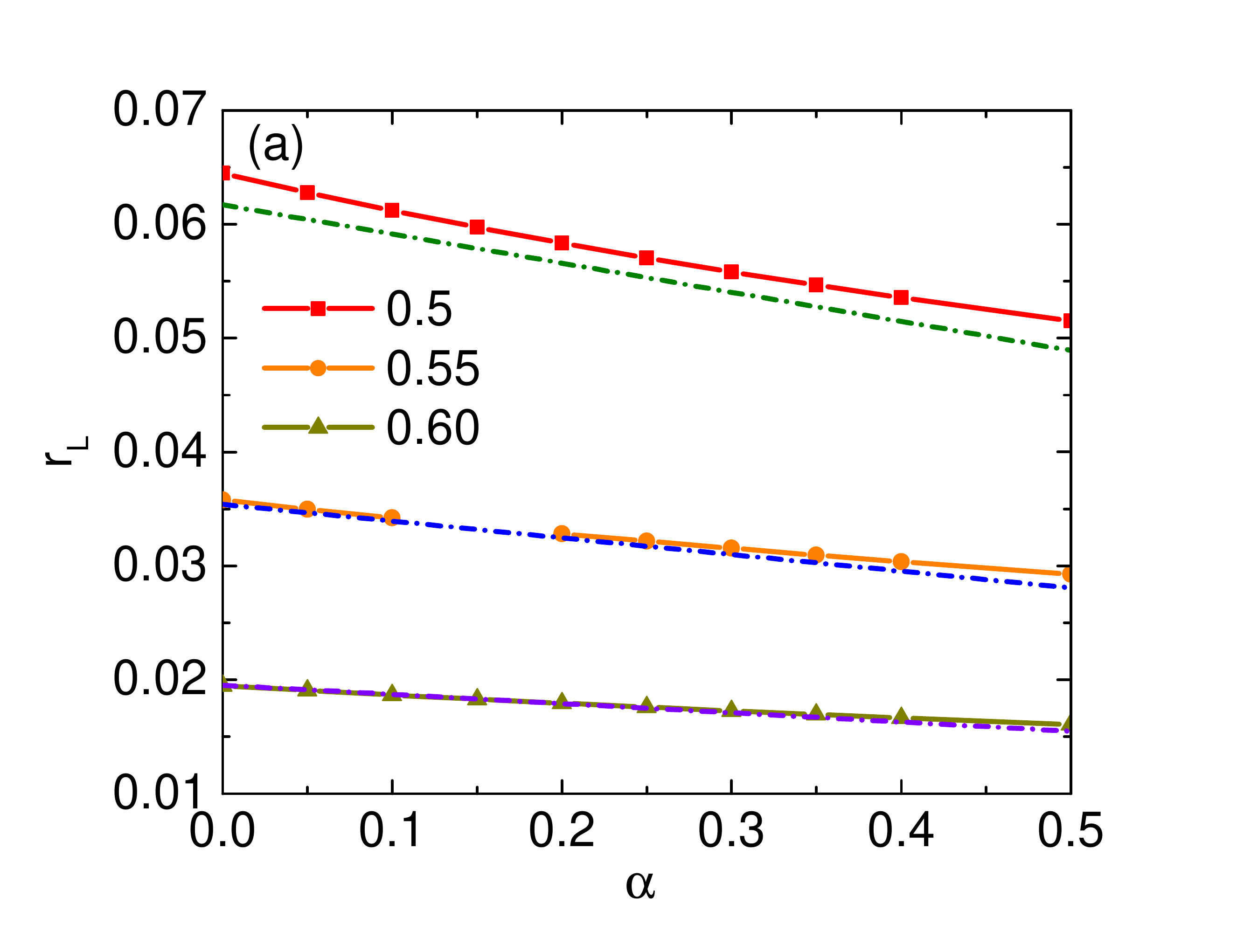}
\includegraphics[width=8cm]{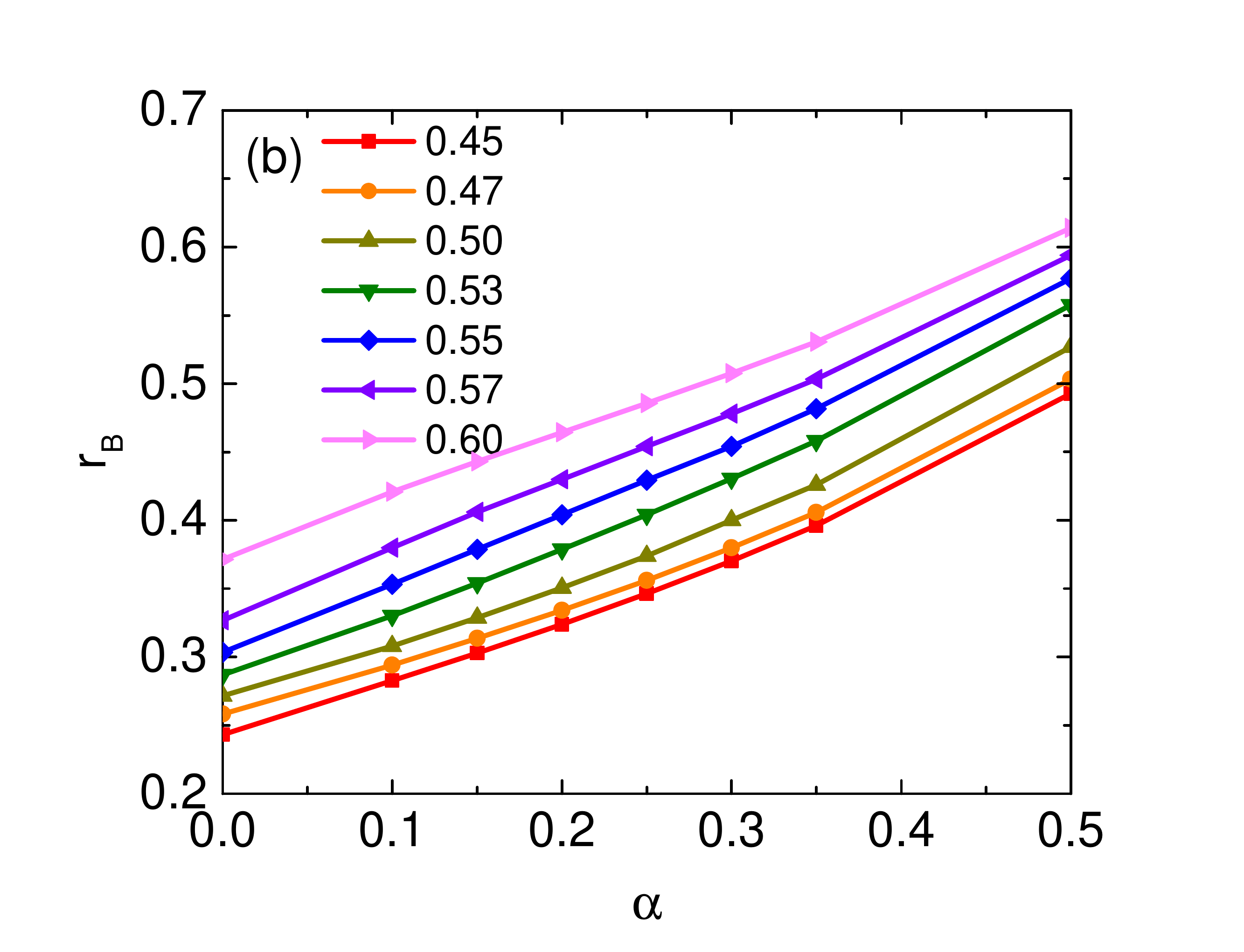}
\caption{\label{fig:3}(Color online) (a) The localization length (units of particle diameter) as a function of pinning fraction at the indicated different volume fractions, $\Phi$ . The solid and dashed-dotted curves corresponds to the full numerical calculations and the ultra-local analytic expression discussed in the text, respectively. (b) The corresponding barrier position as a function of pinning fraction at various volume fractions.}
\end{figure}

Figures \ref{fig:3}a and \ref{fig:3}b show results for the transient localization length and barrier location, respectively, as a function of pinning fraction at high packing fractions. For unpinned systems ($\alpha=0$), the localization length (barrier position) decreases (increases) with volume fraction, and these trends persist at nonzero degrees of pinning. At fixed volume fraction, pinning reduces the localization length in a roughly linear manner. On the other hand, the barrier location, $r_B$, increases with pinning fraction. We physically interpret this trend in the context of 1-component NLE theory which predicts $r_B$
increases with density \cite{10,33}. The increase here with pinning fraction is suggested to be a consequence of a reduced number of pathways for a mobile particle to hop as the cage becomes more rigid and confining.
 
\begin{figure}[htp]
\center
\includegraphics[width=8cm]{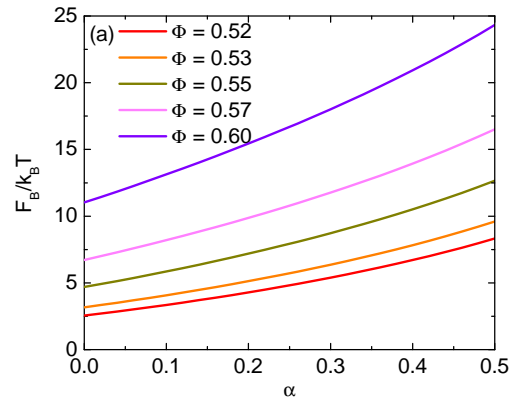}
\includegraphics[width=8cm]{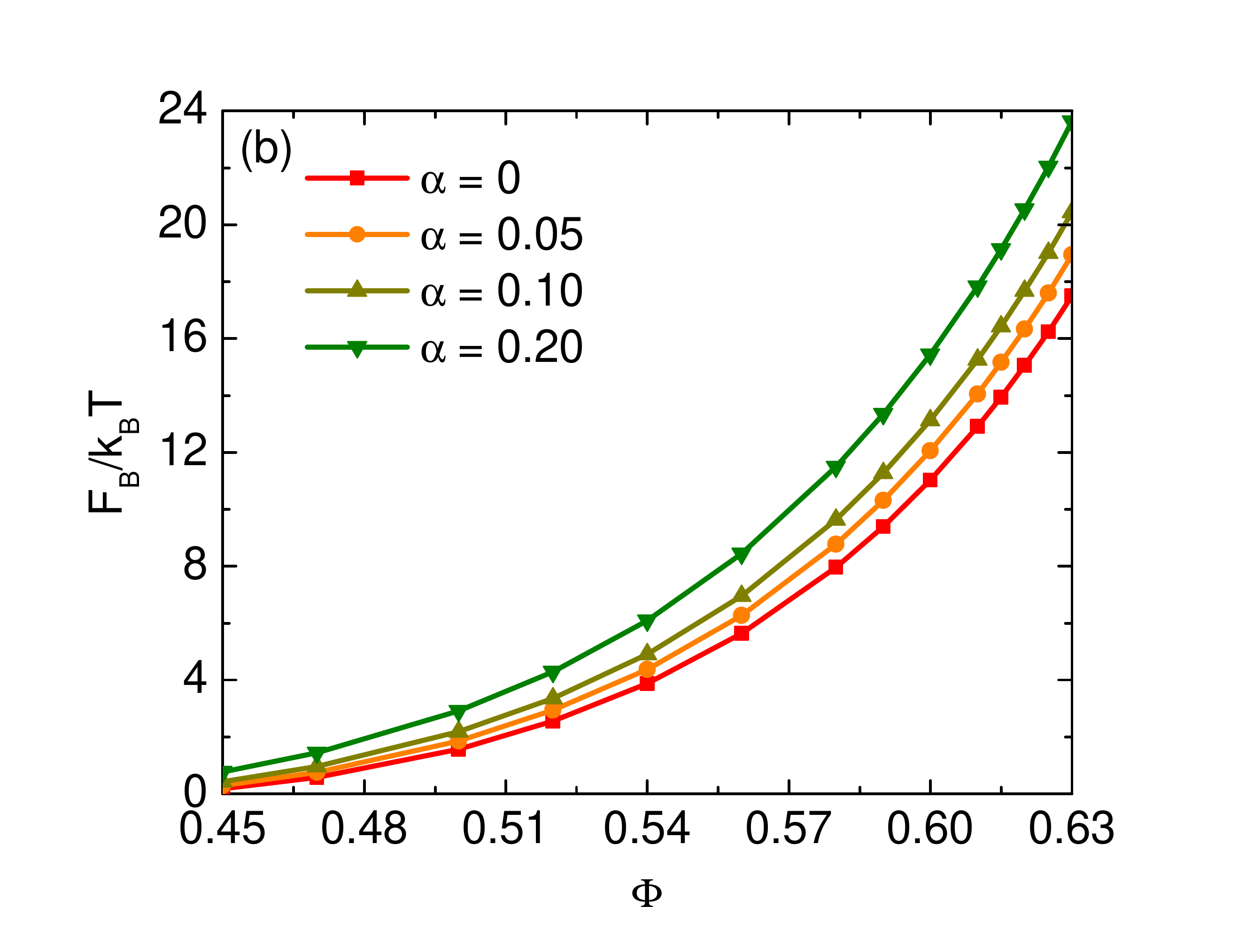}
\caption{\label{fig:4}(Color online) The local cage barrier (in thermal energy units) as a function of (a) pinning fraction at various volume fractions, (b) volume fraction at various pinning fractions.}
\end{figure}

Figure \ref{fig:4} shows the variation of the local barrier height with pinning fraction at fixed volume fraction, and as a function of volume fraction at fixed degree of pinning. As expected, pinning always increases the barrier for local hopping. Figure \ref{fig:4}a shows that the $\alpha$-dependence of the barrier is weakly supra-linear. Qualitatively, the hopping time is proportional to $e^{\beta F_B}$, and thus one expects the relaxation time will grow roughly exponentially with pinning fraction. Figure \ref{fig:4}b shows that the volume fraction dependence of the barrier is relatively modestly enhanced with pinning. In pure hard sphere fluids, the local NLE barrier grows nearly linearly with inverse localization length \cite{7,33,45}. Figure 5 shows that this behavior continues to hold rather well in the presence of pinning, although there are second order deviations. 

Figure \ref{fig:6} uses Eq.(\ref{eq:shear}) plus the localization length results of Fig. \ref{fig:3} to compute how the dynamic shear modulus of the ideal arrested mobile sub-system changes with pinning fraction and volume fraction. Recall there is no change of local structure, and pinned particles enter the calculation only via their effect on the mobile subsystem. The latter enters Eq.(\ref{eq:shear}) via the prefactor ($1-\alpha$) which multiplies density, and  is the leading cause of $G$ decreasing with pinning fraction as seen in Fig.\ref{fig:6}. If this factor is removed, $G$ grows with pinning fraction since mobile particles are more localized. In any case, changes of $G$ with pinning fraction are modest.

\begin{figure}[htp]
\center
\includegraphics[width=8cm]{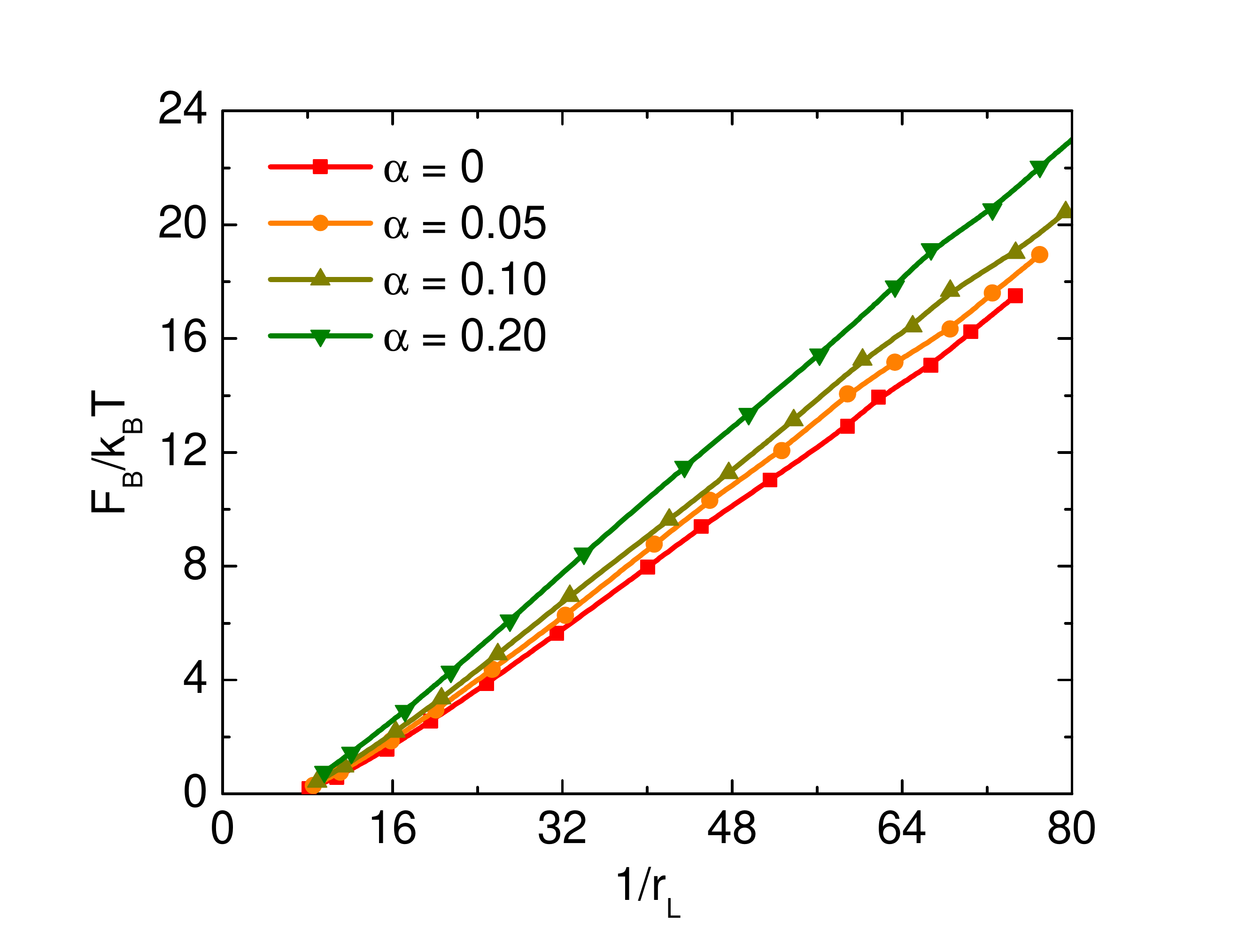}
\caption{\label{fig:5}(Color online) The local cage barrier for various volume fractions as a function of dimensionless inverse localization length, $d/r_L(\alpha,\Phi)$, for a range of volume fractions at 4 fixed values of pinning fraction.}
\end{figure}

\begin{figure}[htp]
\center
\includegraphics[width=8cm]{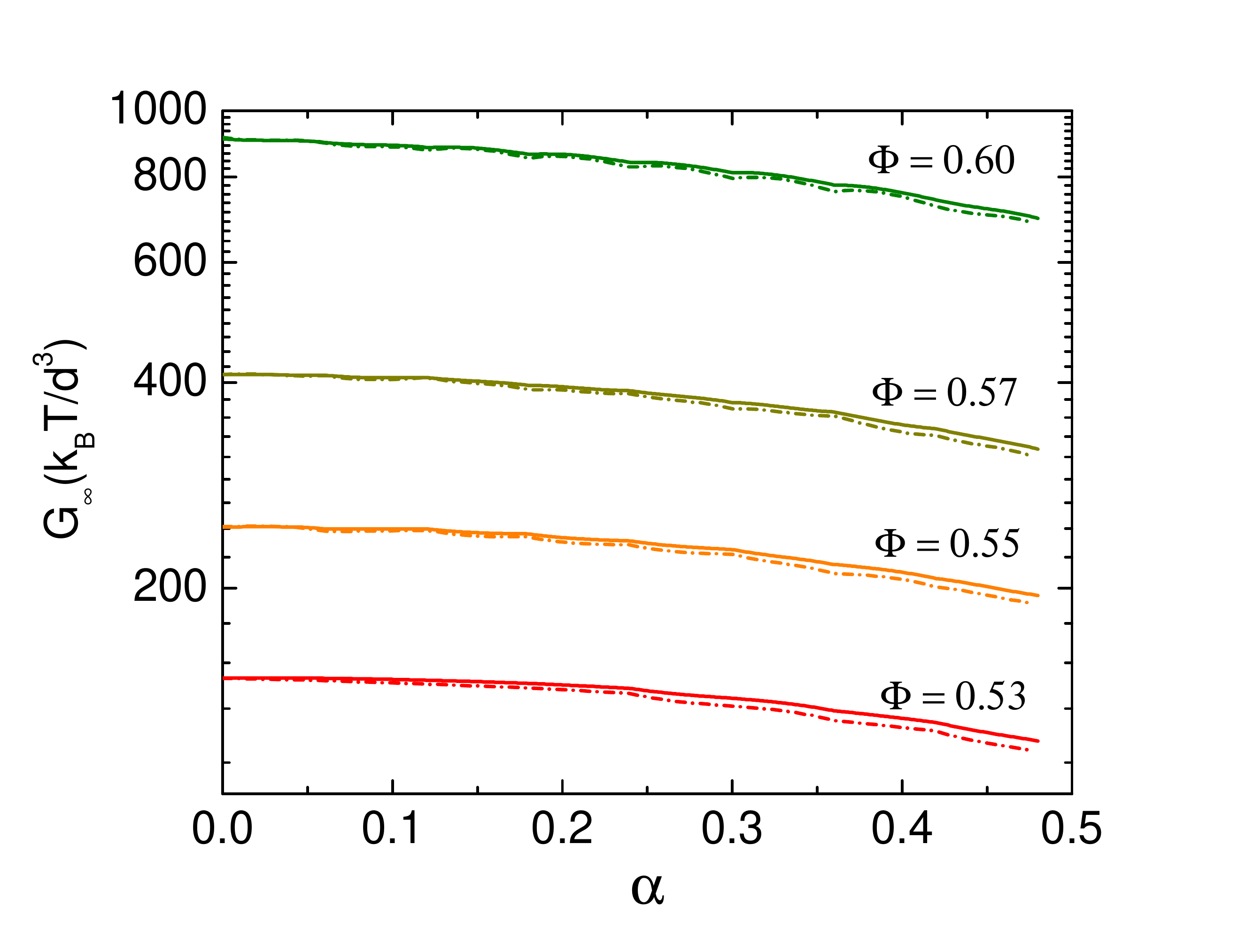}
\caption{\label{fig:6}(Color online) The logarithm of the shear modulus (in units of $k_BT/d^3$) as a function of pinning fraction for various volume fractions. The solid and dashed-dotted curves correspond to the full numerical results and the analytic ultra-local analysis expression discussed in the text, respectively.}
\end{figure}

\subsection{Analytic Analysis}
For hard sphere fluids with barriers beyond a few $k_BT$, much insight has been gained within the NLE framework based on the "ultra-local" analytic analysis \cite{45}. The latter is enabled by high wavevector dominance in the dynamic force vertex of Eqs.(\ref{eq:rL}) and (\ref{eq:1}) and the known analytic form of $c(q)$ in this regime. We do not repeat published technical details \cite{45}.

The critical result of the ultra-local analysis is that for pure hard sphere fluids a single “coupling constant” controls, to leading order, all aspects of the dynamic free energy \cite{45}:
\begin{eqnarray}
\lambda = \Phi g(d)^2,
\end{eqnarray}
where $g(d)$ is the contact value of the pair correlation function. The dynamic vertex in NMCT and NLE theories is related to an effective mean square force experienced by a tagged particle due to its environment, which is dominated by its caging neighbors for short range interactions. This leads to an intuitive result since the “effective force” for hard spheres is an impulse that acts only when particles are in contact, and hence $\sim k_BTg(d)/d$. The contact value is directly related to the thermodynamic dimensionless pressure $P$ (compressibility factor, $Z$) via an exact theorem \cite{7,35}:
\begin{eqnarray}
\Phi g(d) \varpropto Z-1, \quad Z = \frac{\beta P}{\rho}\rightarrow \lambda \varpropto \frac{(Z-1)^2}{\Phi}.
\end{eqnarray}
Prior analytic analysis found ($d/r_L$) $ \varpropto \beta F_B \varpropto \lambda$, relations which connect short time (localization length) and long time (barrier hopping) dynamics, a hallmark of NLE theory \cite{45}.

Ultra-local analytic analysis has been performed for the pinned-mobile system; Appendix C provides some details. The localization length and barrier position follow from the self-consistent equation:
\begin{eqnarray}
\frac{\sqrt{3\pi}}{4}\frac{d}{r_{L,B}} &=& \Phi g^2(d)\left[\sqrt{2}\alpha erfc\left(\frac{q_cr_{L,B}}{\sqrt{6}}\right)\right.\nonumber\\
&+& \left. (1-\alpha)erfc\left(\frac{q_cr_{L,B}}{\sqrt{3}}\right) \right], 
\label{eq:48}
\end{eqnarray}
where $q_c = 2\pi/r_{cage}$ is the lower wavevector cutoff \cite{45}. Since $q_cr_L(\alpha) \ll 1$, one can take $q_cr_L(\alpha) =0$ and obtain for the localization length:
\begin{eqnarray}
r_L(\alpha) &\approx& \frac{\sqrt{3\pi}}{4\Phi g^2(d)}\left[1- (\sqrt{2}-1)\alpha \right] \nonumber\\
&=& r_L(0)\left[1- (\sqrt{2}-1)\alpha \right].
\label{eq:49}
\end{eqnarray}
The predicted linear dependence on pinning fraction is shown in Fig.\ref{fig:3}a, and is in excellent accord with the full numerical calculations. If $q_cr_B$ is sufficiently large then $erfc(x)\approx e^{-x^2}/(x\sqrt{\pi})$ in Eq.(\ref{eq:48}), which allows one to obtain:
\begin{eqnarray}
r_B(\alpha) &=& \frac{1}{q_c}\sqrt{6\ln\left( \frac{4\Phi g^2(d)q_c}{\pi d} \left[\alpha + \sqrt{\alpha^2+\frac{(1-\alpha)\pi d}{4\Phi g^2(d)q_c}} \right]\right)}\nonumber\\
&\approx& \frac{1}{q_c}\sqrt{3\ln\left( \frac{4\Phi g^2(d)q_c}{\pi d}\right)},
\label{eq:50}
\end{eqnarray}
where the final expression follows for large enough values of $\alpha$. Eq.(\ref{eq:50}) captures the increase of the barrier location (and hence jump distance) with pinning fraction, and all the other trends in Fig.\ref{fig:3}b, though not with quantitative accuracy (not shown).

The local barrier height can also be analytically calculated in the ultra-local limit as:
\begin{widetext}
\begin{eqnarray}
\frac{F_B}{k_BT} &=& -3\ln\frac{r_B}{r_L}-4d^2\rho g^2(d)\int_{q_c}^\infty \frac{dq}{2q^2} \left[2\alpha\left(e^{-q^2r_B^2/6}-e^{-q^2r_L^2/6}\right) + (1-\alpha)\left(e^{-q^2r_B^2/3} -e^{-q^2r_L^2/3} \right)\right] \nonumber\\
&=& -3\ln\frac{r_B}{r_L}+\frac{12\Phi g^2(d)}{\sqrt{\pi} q_cd}\left[(1-\alpha)\left(\frac{q_cr_B}{\sqrt{3}}erfc\left(\frac{q_cr_B}{\sqrt{3}} \right)+\frac{\left(e^{-q_c^2r_L^2/3}-e^{-q_c^2r_B^2/3} \right)}{\sqrt{\pi}} \right)\right. \nonumber\\
&+& \left. 2\alpha\left(\frac{q_cr_B}{\sqrt{6}}erfc\left(\frac{q_cr_B}{\sqrt{6}} \right)+\frac{\left(e^{-q_c^2r_L^2/6}-e^{-q_c^2r_B^2/6} \right)}{\sqrt{\pi}}  \right) \right].
\label{eq:52}
\end{eqnarray}
\end{widetext}
It depends on $r_L$, $r_B$, $\Phi$, $g(d)$ and $\alpha$. Further simplification follows by adopting the inequalities  $q_cr_L\ll \sqrt{6}$ and $q_cr_B > \sqrt{6}$ (reasonable in the high barrier regime), yielding:
\begin{eqnarray}
\frac{F_B}{k_BT} = -3\ln\frac{r_B}{r_L}+\frac{12\Phi g^2(d)}{\pi q_cd}\left(1+\alpha\right)+3
\label{eq:53}
\end{eqnarray}
In practice, the second term is more dominant. Eq.(\ref{eq:53}) is consistent with the trends in Fig.4a, including the roughly linear growth of $F_B$ with pinning fraction at fixed density and the near linear proportionality between the barrier height and inverse localization length seen in Fig.\ref{fig:5}.

An analytic analysis of the dynamic shear modulus can be straightforwardly performed based on Eq.(\ref{eq:shear}). One obtains:
\begin{eqnarray}
G(\alpha) &\approx& \frac{9\Phi k_BT(1-\alpha)}{5\pi d r_L^2(\alpha)} \approx G(0)\left[1- \left(3-2\sqrt{2}\right)\alpha \right. \nonumber\\
 &-& \left.\left(4\sqrt{2}-5\right)\alpha^2 - \left(3-2\sqrt{2}\right)\alpha^3\right].
\label{eq:shearModulus}
\end{eqnarray}
It is inversely proportional to the localization length squared, or equivalently one power of the harmonic spring constant of the dynamic free energy, $K_0$. Figure \ref{fig:6} shows good agreement between the analytic and numerical results. Eqs. (\ref{eq:53}) and (\ref{eq:shearModulus}) also imply the inter-relations:
\begin{eqnarray}
F_B(\alpha)&\varpropto&\sqrt{G(\alpha)}, \\
\quad F_B(\alpha)r_L(\alpha) &\varpropto& (1+\alpha)/(1+(\sqrt{2}-1)\alpha).
\label{eq:25}
\end{eqnarray}
Eq.(\ref{eq:25}) explains the secondary trend in Fig. \ref{fig:5} that increasing $\alpha$ increases the barrier at fixed $1/r_L$.
\subsection{Mean Hopping Time Results}
We now consider the mean barrier hopping time, taken as a surrogate for average alpha relaxation time $\tau_\alpha$. It follows from the Kramers mean first passage time as \cite{7,46}:
\begin{eqnarray}
\tau_\alpha \approx \tau_s\frac{2\pi}{\sqrt{K_0K_B}}e^{\beta F_B}.
\label{eq:26}
\end{eqnarray}
Eq.(\ref{eq:26}) applies when the barrier is beyond several thermal energy units, and here $\tau_s$ is the short length/time scale dynamical process associated with cage-corrected binary (Enskog) collisions \cite{7,33}. We assume the latter is unaffected by pinning. Any modification of $\tau_s$ by pinning is a small effect given it enters Eq.(\ref{eq:26}) as a prefactor. Thus, $\tau_s$ is given by the prior employed hard sphere fluid expression (for Newtonian dynamics) \cite{7}:
\begin{eqnarray}
\tau_s &=& \tau_0\left[1+\frac{1}{36\pi\Phi}\int_{0}^{\infty}dq \frac{q^2\left[ S(q)-1 \right]^2}{S(q)+b(q)} \right], \nonumber\\
\tau_0 &=& \frac{g(d)}{24\rho d^2}\sqrt{\frac{M}{\pi k_BT}},
\label{eq:short}
\end{eqnarray}
where $b^{-1}(q) = 1-j_0(q)+2j_2(q)$, $j_n(x)$ is the spherical Bessel function of order $n$, $M$ is the particle mass, and $\tau_0$ is a "bare" Boltzmann-like time scale relevant to the low density limit.

\begin{figure}[htp]
\center
\includegraphics[width=8cm]{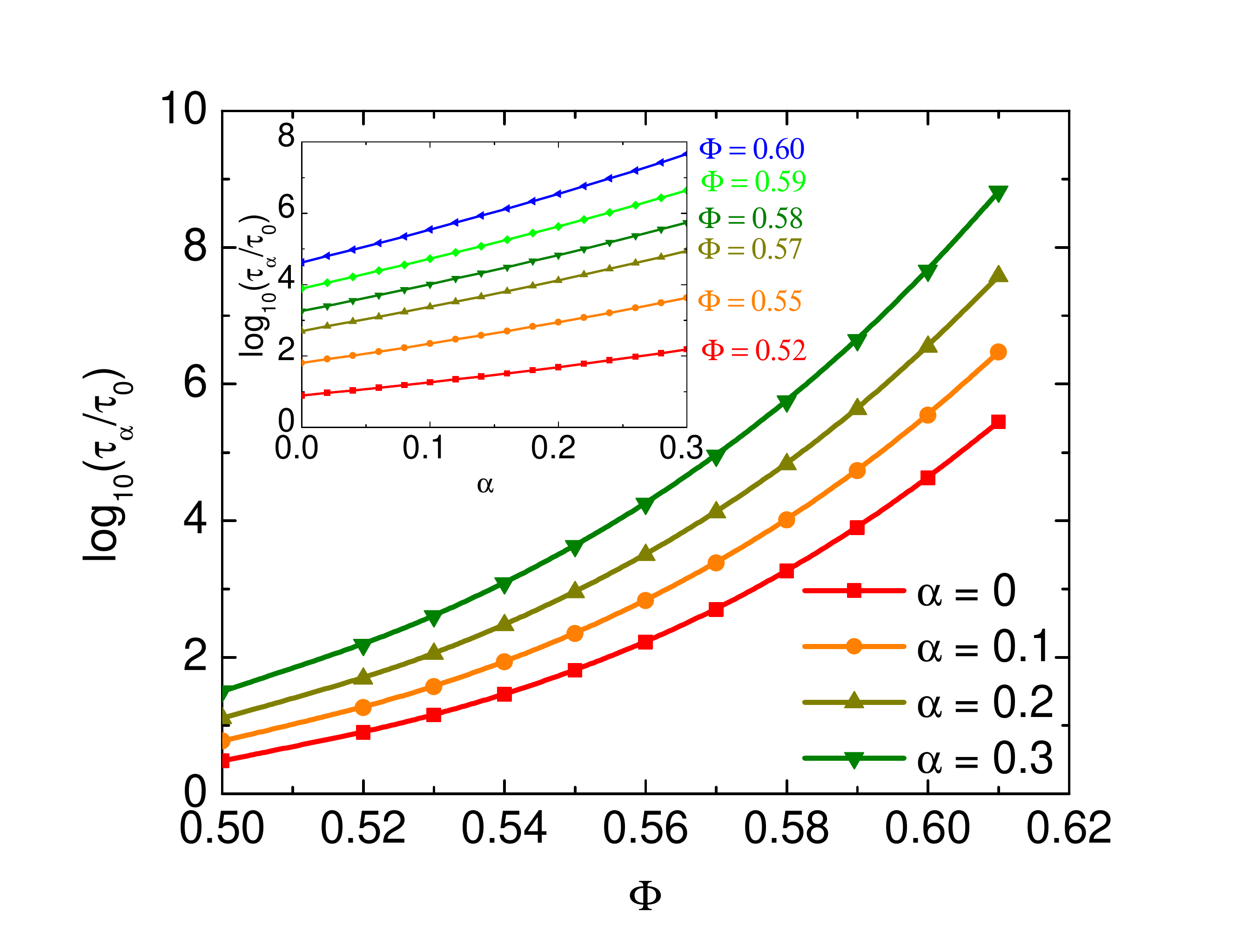}
\caption{\label{fig:7}(Color online) Log-linear plot of the dimensionless mean barrier hopping time computed using NLE theory versus volume fraction for several pinning fractions, and (inset) versus pinning fraction for several volume fractions.}
\end{figure}

Figure \ref{fig:7} presents NLE theory calculations of the alpha time. The main frame shows its volume fraction dependence becomes stronger as pinning fraction increases. This is largely due to the higher barrier per Fig.\ref{fig:4}. The inset of Fig. \ref{fig:7} shows the alpha time at fixed volume fraction grows in a weakly supra-exponential manner with pinning fraction, as expected based on Fig.\ref{fig:4}. Over the range of pinning fractions probed in simulations (up to $\alpha\sim0.15$), the nearly exponential growth with $\alpha$ occurs with a slope that grows with increasing volume fraction. Based on the inverse temperature volume fraction correspondence \cite{3,40}, this trend is consistent with simulations in this regime (up to $\alpha\sim0.15$) at high and intermediate temperatures above the empirical MCT value.

\begin{figure}[htp]
\center
\includegraphics[width=8cm]{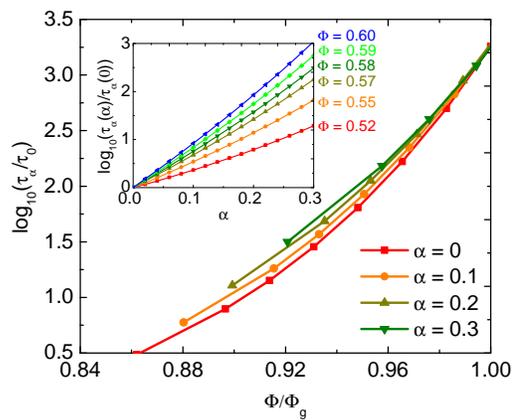}
\caption{\label{fig:8}(Color online) Normalized Angell-like plot of the dimensionless NLE theory alpha relaxation time versus scaled volume fraction, $\Phi/\Phi_g$, for various random pinning fractions. Inset- Log-linear plot of the alpha time normalized by its pure fluid analog as a function of pinning fraction at several volume fractions.}
\end{figure}

Figure \ref{fig:8} presents the relaxation time calculations in two distinct normalized formats. The inset shows how it grows with pinning fraction relative to the volume-fraction-dependent pure fluid analog. An exponential growth law is clearly seen, along with only of order one decade enhancement at a pinning fraction of 15 $\%$ even at a high volume fraction of $\sim 0.58$ (empirical MCT crossover). The main frame shows the analog of an Angell plot where volume fraction is scaled by its value where the alpha time of pinned systems equal its unpinned fluid analog at $\Phi = 0.58$. This procedure operationally defines a kinetic vitrification volume fraction, $\Phi_g$, which decreases with pinning fraction (see Fig.2). The theory predicts dynamic fragility weakly decreases with pinning, as evidence by the weaker density variation in Fig. \ref{fig:8}, a trend in  qualitative accord with simulations of thermal liquid models \cite{25,28}. 

\subsection{NLE Theory versus Simulation}
We recall that bulk (no pinning) colloid experiments and hard sphere fluid simulations typically probe only roughly 3 decades in relaxation time in the “glassy precursor regime” spanning the range of $\Phi\sim 0.5-0.58$ \cite{1,3}. For this initial slowing down regime, NLE predicts the alpha time grows by a smaller amount of order 1.5 decades. Hence, collective elastic effects seem already important. At the even higher volume fractions probed in more recent simulation and experimental work \cite{47}, NLE theory was found to strongly under predict the alpha time \cite{7,8,30,48}. Hence, one might anticipate NLE theory (strongly) under predicts the effect of pinning on relaxation. As discussed below, this is what we find. The one caveat, which we believe is a major one, is whether quantitative or subtle trends deduced based on isochoric simulations that lower temperature can be expected to present in our isothermal results for the effect of pinning as a function of density. We are unaware of simulations that have definitively addressed this question. Our intuition is there could be major differences.

Near the empirical MCT crossover of bulk ECNLE theory ($\Phi\sim 0.58-0.59$), NLE theory predicts only roughly 1 decade of slowing down at $\alpha\sim 0.15$ compared to the alpha time of the pure system. In contrast, simulations of a binary mixture of soft repulsive harmonic spheres \cite{23,25} over a temperature range where the bulk alpha time grows by 5-6 decades find exponential enhancements of the alpha time with pinning fraction which reach a factor of $\sim 10000$ at $\alpha\sim 0.15$ near the empirical MCT temperature.  Simulations of binary LJ mixtures \cite{26} find a weakly supra-exponential growth of the alpha time with pinning fraction which is enhanced with cooling, reaching a factor of $\sim 1000$ at $\alpha\sim 0.15$ for $T/T_{MCT,empirical}\approx 1.3$. Studies of other binary soft sphere mixtures \cite{27} up to $\alpha\sim 0.1$ over a modest range of temperature (bulk alpha time grows by 2 decades) find an exponential growth of the relaxation time with $\alpha$ by a factor of $\sim 1000$ at the lowest $T$ studied. Simulations of yet other 2d and 3d model mixtures \cite{28} find similar trends up to $\alpha\sim 0.1$, but the alpha time grows significantly more strongly than exponential with pinning fraction. Thus, in the glassy precursor regime probed in diverse simulations, although there are quantitative variations, the qualitative trends are broadly similar including a roughly exponential growth of time scale with pinning fraction. This trend is captured by NLE theory but with a magnitude strongly under-predicted. Our hypothesis is that collective elastic effects are important even in the dynamic precursor regime, a natural deduction given the known situation for bulk fluids \cite{7,8}.

We note one qualitative deviation between the NLE theory and repulsive harmonic sphere simulations [25] which appears to have probed the lowest temperatures to date. They found the relative increase of the relaxation time with pinning fraction, which grows with cooling at relatively high and intermediate temperatures, slows down and appears to saturate near the empirical MCT temperature. This trend is seemingly in contrast to the NLE theory result that the relative growth monotonically increases with density. We do believe the latter trend is correct for the purely \emph{local} physics that NLE theory addresses. Curiously, other simulations \cite{23,26,27,28} do not report the aforementioned behavior, and whether the reason is they use different interparticle potentials and/or do not probe to as effectively low temperature is unclear to us.

\section{Collective Elastic Effects in Pinned-Mobile Systems}
\subsection{Qualitative Considerations}
The discussion in section IIIE raises two fundamental theoretical questions. For hard spheres, will the proper generalization of ECNLE theory that includes collective elastic effects in the pinned-mobile hard sphere system predict a non-monotonic variation or saturation-like behavior of the alpha time at relatively high volume fractions?  Should such a feature even be present if density is the control variable versus temperature under constant volume conditions? We have no answer to the second question, and suggest new simulation studies are necessary. For the former question, we first offer a qualitative discussion of how random pinning might change the collective elastic barrier which involves multiple distinct physical effects that may be affected differently by pinning. The full problem, including possible pinning-induced strain field localization, is presently under study. Section IVB presents our initial effective medium analysis.

The collective elastic barrier involves three key contributions in Eq.(\ref{eq:elastic2}) (see Fig.\ref{fig:1} for a schematic) \cite{7}. (1) The microscopic particle jump distance which sets the amplitude of cage dilation and hence the required elastic displacement field fluctuation. (2) The degree of transient particle localization ($r_L$) or harmonic spring constant ($K_0$), which sets the energy scale for elasticity and the collective elastic barrier. (3) The spatial form of the strain or displacement field as a function of distance from the cage center. Contributions (1) and (2) are local properties determined by NLE theory, and they both change in the direction of a larger elastic barrier for all volume fractions as pinning fraction increases. Issue (3) is complex since it requires knowing how the excluded volume associated with quenched disorder (immobile particles) modifies the facilitating elastic strain field. Physically we expect the latter may become spatially localized since the randomly pinned particles cannot move and rigorously expel it. If true, such displacement field localization presumably reduces the elastic barrier, and increasingly so as more particles are pinned. Hence, whether more pinning increases or decreases the collective elastic barrier would seem to be a subtle problem that depends on three competing factors. A possible scenario for the pinning enhancement of the relaxation time to stop growing at high enough density (or low enough temperature) is that point (3) becomes dominant, a perhaps plausible speculation if the strain field becomes exponentially localized in space.  However, the problem seems even more subtle since, as argued in the literature \cite{25}, under sufficiently deep supercooling conditions that can be probed in the laboratory, pinning enhancement of the relaxation time is expected to again become stronger with cooling or densification; this regime is presently beyond the capability of computer simulation. 
\subsection{{Na\"ive} Effective Medium Approximation}
In bulk globally homogeneous fluids, ECNLE theory adopts an elastic continuum model as the technical tool to determine the spatial form of radially-symmetric displacement field outside the cage based on solving \cite{11}: 
\begin{eqnarray}
\left(\mathcal{K}_B+\frac{G}{3}\right)\nabla(\nabla .\mathbf{u}) + G\nabla^2\mathbf{u}  = 0,
\label{eq:40}
\end{eqnarray}
where $\mathcal{K}_B$ and $G$ are the bulk and dynamic shear modulus, respectively, and $\mathbf{u}$ is the vector of the displacement field. Randomly pinning particles introduces quenched spatial disorder, fluctuations in local mechanical stiffness, and the hard constraint that the mobile particle facilitating displacement field cannot penetrate the finite excluded volume presented by the pinned particles. How to determine the modified strain field is an open problem. Here, we analyze only the simplest approximation.

Recall our simple treatment of the local cage scale (NLE theory) aspect whence particle pinning only enters via setting their Debye-Waller factors to unity, from which we compute their effect on all key properties of the dynamic free energy. This analysis seems akin to the simplest effective medium approach, and we adopt a similar perspective for the elastic barrier. The dynamic free energy predicts the required changes with pinning of the localization well curvature and jump distance in Eq.(\ref{eq:elastic2}), and we make the strong assumption of effective spatial homogeneity and use the unpinned form of the strain field spatial dependence in Eq.(\ref{eq:elastic1}). This may over-predict the spatial range of the displacement field and the effect of pinning on the elastic barrier as discussed in section IIIE. 
\subsection{Numerical Results}
\begin{figure}[htp]
\center
\includegraphics[width=8cm]{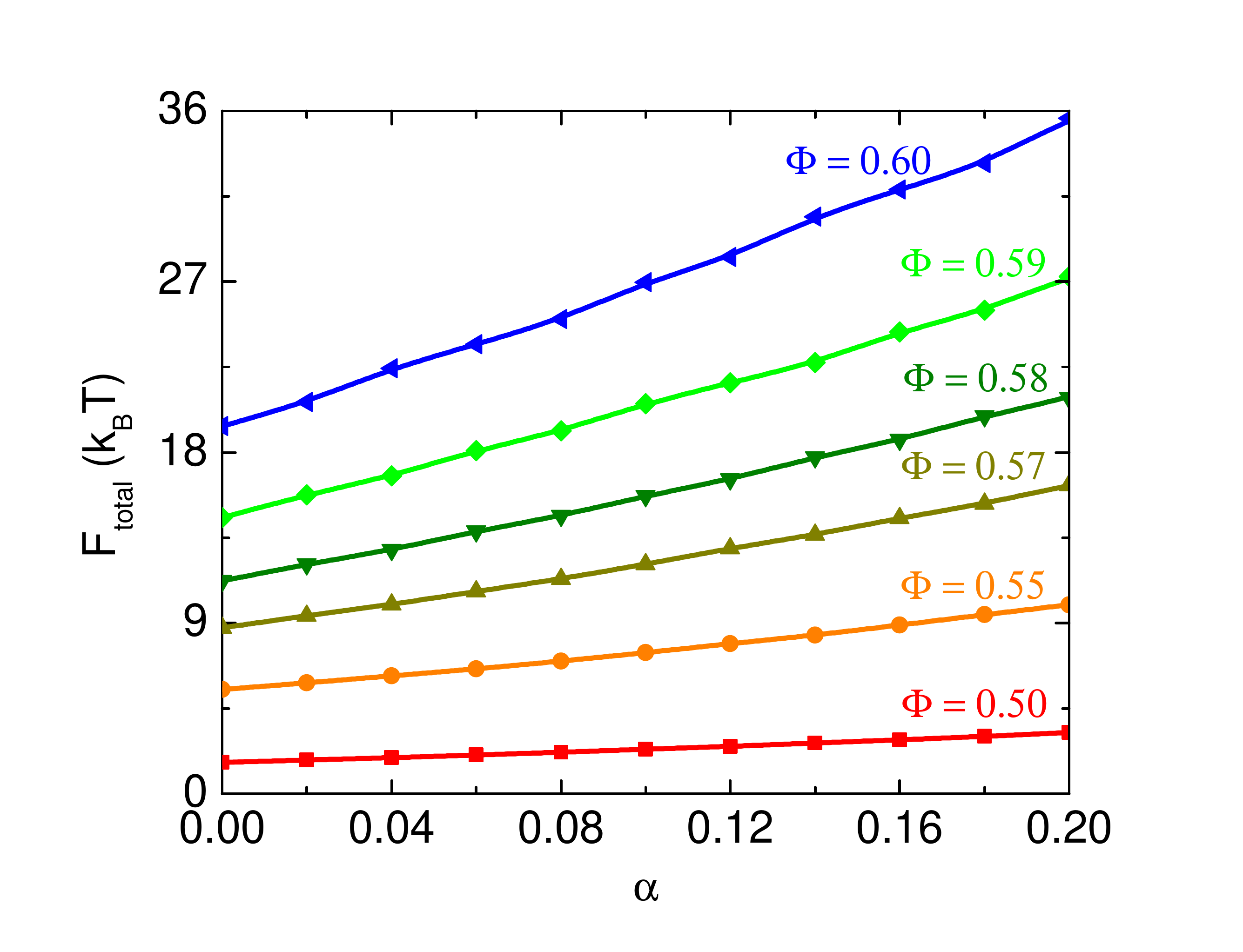}
\caption{\label{fig:9}(Color online) The total (local cage plus collective elastic) barrier as a function of pinning fraction at several volume fractions.}
\end{figure}

Figure \ref{fig:9} shows ECNLE theory calculations of the \emph{total} barrier. Remarkably, the nearly linear growth with pinning fraction is again found. This implies that, qualitatively, the alpha relaxation time grows exponentially (or weakly supra-exponentially) with pinning fraction, and with a slope that grows monotonically with volume fraction.

\begin{figure}[htp]
\center
\includegraphics[width=8cm]{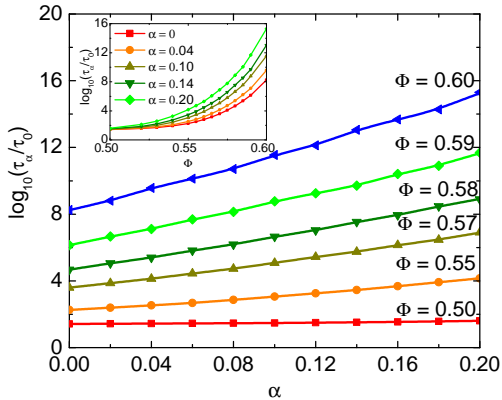}
\caption{\label{fig:10}(Color online) Log-linear plot of the dimensionless mean alpha relaxation time calculated using ECNLE theory versus the fraction of pinned particles for several volume fractions. Inset – Analogous results plotted versus volume fraction at several pinning fractions.
}
\end{figure}

The corresponding hopping times are shown in Fig.\ref{fig:10} based on the previously developed expression for the alpha relaxation time \cite{7,8}:
\begin{eqnarray}
\tau_\alpha \approx \tau_s\left[1+\frac{2\pi}{\sqrt{K_0K_B}}e^{\beta(F_B+F_e)} \right].
\label{eq:41}
\end{eqnarray}
The theory predicts the alpha time near the empirical MCT crossover volume fraction of $\sim 0.58-0.59$ is $\sim 3$ decades larger at a pinning fraction of 15$\%$. As discussed in section III.E, this is a reasonable value compared to various simulation studies. The magnitude of the alpha time increase with pinning fraction monotonically grows with volume fraction. This seems intuitive to us, but conflicts with one simulation \cite{25} which found this dependence saturates at low enough temperatures approaching the empirical MCT value. We recall the subtle issue that simulations which vary temperature at fixed density may be quite different for some of the pinning physics than for hard spheres where slower relaxation and barriers are induced by increasing density. An Angell-like plot of the ECNLE theory results of Fig.10 that is analogous to Fig. \ref{fig:8} again shows that the dynamic fragility decreases with pinning fraction (not shown).

\section{Conclusions}
We have extended the microscopic NLE theory for the local cage scale single particle activated dynamics in bulk liquids to treat the effect of random pinning under the neutral confinement condition. The theory was analyzed and implemented for hard spheres. As the pinned fraction grows, all aspects of local cage confinement as quantified by the dynamic free energy are enhanced: the localization length of mobile particles decreases modestly, while the barrier location and hence jump distance grow substantially. The local barrier increases in a weakly supra-linear manner, resulting in a weakly supra-exponential growth of the mean alpha time with pinning fraction. The effect of pinning on the barrier and relaxation time grows with volume fraction. Analytic analysis in the so-called "ultra-local limit" was performed for the pinned-mobile system. The derived results agree well with our numerical results, and provide additional insight concerning the numerically obtained trends.

Collective elastic fluctuations are of critical importance when barriers become substantial. They were analyzed by extending the homogeneous fluid ECNLE theory to the pinned-mobile system based on the simplest effective medium approximation. Pinning is then predicted to monotonically enhance the elastic barrier, and more so at higher volume fraction. Changes of the relaxation time due to pinning become order(s) of magnitude larger than predicted by the local NLE approach.

The present theory is easily extended to treat any spherical particle model such as soft repulsive spheres or WCA fluids. Additional complications such as vibrating pinned particles or attractive interactions between the pinned and mobile particles can be straightforwardly treated. The former reduces the effect of random pinning on dynamical slowing down, while the latter is expected to enhance it and is especially relevant for using the random pinning model as a crude mimic of real porous materials.

The theoretical results for hard spheres were qualitatively and semi-quantitatively contrasted with simulations of spherical particle thermal liquids. Similarities were identified in the dynamic precursor regime, including a roughly exponential, or weakly supra-exponential, growth of the alpha time and reduced fragility with pinning fraction. However, large quantitative deviations between the NLE theory results and simulations emerge corresponding to strong under predictions of the extent that pinning increases the relaxation time. The {na\"ive} extension of ECNLE theory to the pinned particle system appears to correct this aspect, yielding exponential growth of the alpha time with reasonable magnitudes. This enhancement monotonically grows with volume fraction. Thus, the tendency of pinning effects on the relaxation time to slow down or even become invariant to temperature under cold enough conditions observed in one simulation study \cite{25} is not captured. Whether this is a reflection of missing physics in the theory, or that how pinning slows motion based on constant volume cooling is different than increasing density isothermally for specific subtle effects, or some other complication, is unclear. 

Future work is aimed at going beyond the {na\"ive} effective medium description of how pinning affects the collective elastic part of the problem. A key missing feature of our present work is that at very high volume fraction or low temperature random pinning of finite excluded volume obstacles may spatially localize the displacement field in a manner that depends on volume fraction and pinning fraction. The construction of a theory for this effect is under development.


\begin{acknowledgements}
This work was performed at the University of Illinois at Urbana-Champaign and was supported by DOE-BES under Grant No. DE-FG02-07ER46471 administered through the Frederick Seitz Materials Research Laboratory.
\end{acknowledgements}



\appendix\markboth{Appendix}{}
\renewcommand{\thesection}{\Alph{section}}
\numberwithin{equation}{section}
\section{Mixture Static and Short Time Dynamic Structure Factors}
To implement the NMCT and NLE theories in Eqs.(\ref{eq:rL}) and (\ref{eq:1}) requires the spherical particle binary mixture (species labels 1,2) direct correlation functions and partial collective structure factors as determined using the Ornstein-Zernike (OZ) matrix integral equations. We simply quote the standard results \cite{34} where $h_{ij}(r) = g_{ij}(r)-1$:

\begin{eqnarray}
S_{ij}(q) & =& \delta_{ij} + \sqrt{\rho_i\rho_j}h_{ij}(q),\nonumber\\
S_{11}(q) &=& \frac{1-\rho_2c_{22}(q)}{[1-\rho_1c_{11}(q)][1-\rho_2c_{22}(q)]-\rho_1\rho_2c_{12}(q)c_{21}(q)},\nonumber\\
S_{21}(q) &=& \frac{\sqrt{\rho_1\rho_2}c_{21}(q)}{[1-\rho_1c_{11}(q)][1-\rho_2c_{22}(q)]-\rho_1\rho_2c_{12}(q)c_{21}(q)},\nonumber\\
S_{12}(q) &=& \frac{\sqrt{\rho_1\rho_2}c_{12}(q)}{[1-\rho_1c_{11}(q)][1-\rho_2c_{22}(q)]-\rho_1\rho_2c_{12}(q)c_{21}(q)},\nonumber\\
S_{22}(q) &=& \frac{1-\rho_1c_{11}(q)}{[1-\rho_1c_{11}(q)][1-\rho_2c_{22}(q)]-\rho_1\rho_2c_{12}(q)c_{21}(q)}.\nonumber\\
\label{eq:61}
\end{eqnarray}
For neutral confinement all direct correlation functions are identical.

Equations (\ref{eq:hydro}) and (\ref{eq:matrix}) define our model for the short time collective partial dynamic structure factors. Because species 2 is a pinned, effectively $\zeta_{s,2} \rightarrow \infty$, and hence $\Omega_{22}(q) = 0$ and $\Omega_{21}(q) = 0$ in Eq.(\ref{eq:matrix}). Using this simplification and $c_{ij}(q)=c(q)$ yields:
\begin{eqnarray}
\Omega_{11}(q) &=& \frac{k_BT}{\zeta_{s,1}}q^2(1-\rho_1C_{11}), \nonumber\\
\Omega_{12}(q) &=& -\frac{k_BT}{\zeta_{s,1}}q^2\rho_1C_{12}, \nonumber\\
\Omega_{av}(q) &=& \frac{\Omega_{11}(q) + \Omega_{22}(q)}{2} = \frac{\Omega_{11}(q)}{2},\nonumber\\ 
\Delta(q) &=& \Omega_{11}(q)\Omega_{22}(q)-\Omega_{12}(q)^2 = 0.
\end{eqnarray}
Solving for the partial dynamic structure factors involves two relaxation modes \cite{34}:
\begin{eqnarray}
S_{11}(q,t) &=& a_Ie^{\Gamma_It}+a_ce^{\Gamma_ct},\nonumber\\ 
S_{21}(q,t) &=& b_Ie^{\Gamma_It}+b_ce^{\Gamma_ct}.
\label{eq:S112}
\end{eqnarray}
Straightforward algebra then yields for the relaxation rates:
\begin{eqnarray}
\Gamma_{I}(q) &=& \Omega_{av}(q)- \sqrt{\Omega_{av}(q)^2-\Delta(q)^2} = 0,\nonumber\\
\Gamma_{c}(q) &=& \Omega_{av}(q)+ \sqrt{\Omega_{av}(q)^2-\Delta(q)^2} = \Omega_{11}(q),\nonumber\\
\end{eqnarray}
and amplitudes:
\begin{eqnarray}
a_I(q) &=& \frac{(\Gamma_I(q)-\Omega_{22}(q))S_{11}(q) + \Omega_{12}(q)S_{21}(q)}{\Gamma_I(q)-\Gamma_c(q)}\nonumber\\ &=& \frac{\rho_1C_{12}S_{21}}{1-\rho_1C_{11}},\nonumber\\
b_I(q) &=& \frac{(\Gamma_I(q)-\Omega_{11}(q))S_{21}(q) + \Omega_{21}(q)S_{11}(q)}{\Gamma_I(q)-\Gamma_c(q)}\nonumber\\ &=& \frac{\rho_1C_{12}S_{21}}{1-\rho_1C_{11}} = S_{21}(q).
\end{eqnarray}

\begin{eqnarray}
a_c(q) &=& \frac{(\Gamma_c(q)-\Omega_{22}(q))S_{11}(q) + \Omega_{12}(q)S_{21}(q)}{\Gamma_c(q)-\Gamma_I(q)}\nonumber\\
 &=& S_{11}-\frac{\rho_1C_{12}S_{21}}{1-\rho_1C_{11}},\nonumber\\
b_c(q) &=& \frac{(\Gamma_c(q)-\Omega_{11}(q))S_{21}(q) + \Omega_{21}(q)S_{11}(q)}{\Gamma_c(q)-\Gamma_I(q)}\nonumber\\ &=& 0.
\end{eqnarray}
Combining all the above, one obtains:
\begin{eqnarray}
S_{11}(q,t) &=&  \frac{\rho_1c(q)S_{21}(q)}{1-\rho_1c(q)} \nonumber\\
&+& \left(S_{11}(q)-\frac{\rho_1c(q)S_{21}(q)}{1-\rho_1c(q)} \right)e^{-D_1q^2(1-\rho_1c(q))t} \nonumber\\
S_{21}(q,t) &=& S_{21}(q).
\label{eq:S113}
\end{eqnarray}
$S_{22}(q)$ and $S_{12}(q)$ follow by interchanging the labels 1 and 2 in the above results to obtain:
\begin{eqnarray}
S_{12}(q,t) &=&  \frac{\rho_1c(q)S_{22}(q)}{1-\rho_1c(q)} \nonumber\\
&+&\left(S_{12}(q) -\frac{\rho_1c(q)S_{22}(q)}{1-\rho_1c(q)}\right)e^{-D_1q^2(1-\rho_1c(q))t}, \nonumber\\
S_{22}(q,t) &=& S_{22}(q). 
\end{eqnarray}

\section{Derivation of NMCT and NLE Theories for the Pinned-Mobile System}
To construct the self-consistent NMCT equations one takes the long time limit of the appropriate GLE’s. For the pinned-mobile system, this is achieved via the same mapping employed for the 1-component system: $6D_{s,1}=6k_BTt/\zeta_{s,1}\rightarrow r_L^2$. Implementing this and using the partial collective dynamic structure factor expressions of Appendix A in Eq.(\ref{eq:rL}) yields:

\begin{widetext}
\begin{eqnarray}
\frac{9}{r^2_{L1}}&=& \int \frac{d\mathbf{q}}{(2\pi)^3}q^2e^{-q^2r_{L1}^2/6}\left[c(q)S_{11}(q,t\rightarrow \infty)c(q) + c(q)S_{21}(q,t\rightarrow \infty)c(q) \right.\nonumber\\
& &\left.  c(q)S_{22}(q,t\rightarrow \infty)c_{21}(q) + c(q)S_{12}(q,t\rightarrow \infty)c(q)\right]\nonumber\\
&=& \int \frac{d\mathbf{q}}{(2\pi)^3}q^2e^{-q^2r_{L1}^2/6}\left[c(q)^2\left(\frac{\rho_1c(q)S_{21}}{1-\rho_1c(q)} + \left(S_{11}-\frac{\rho_1c(q)S_{21}}{1-\rho_1c(q)} \right)e^{-q^2r_{L1}^2(1-\rho_1c(q))/6} \right) + c(q)S_{22}(q)c(q) \right.\nonumber\\
& & \left.  c(q)S_{21}(q)c(q) + c(q)c(q)\left(\frac{\rho_1c(q)S_{22}}{1-\rho_1c(q)} + \left(S_{12}-\frac{\rho_1c(q)S_{22}}{1-\rho_1c(q)} \right)e^{-q^2r_{L1}^2(1-\rho_1c(q))/6} \right) \right]\nonumber\\
&=& \int \frac{d\mathbf{q}}{(2\pi)^3}q^2e^{-q^2r_{L1}^2/6}\left[A+Be^{-q^2r_{L1}^2(1-\rho_1c(q))/6} \right]
\label{eq:appendix1}
\end{eqnarray}
\end{widetext}
In the final equality, the factors A and B are defined. After major simplifications using the equilibrium relations of OZ mixture theory, these factors are given by:

\begin{eqnarray}
A &=&  \frac{c(q)S_{12}}{\rho_1\left(1-\rho_1c(q) \right)}, \nonumber\\
B &=& \frac{\rho_1c(q)^2}{1-\rho_1c(q)}.
\end{eqnarray}

Employing the above results yields Eq.(14) of the main text. The corresponding dynamic free energy follows as in prior work for 1-component systems \cite{7,10}, thereby yielding Eq.(\ref{eq:1}).

\section{Ultra-Local Analytic Analysis}
The analytic results presented in section III.C are derived in precisely the same way discussed in detail previously for the 1-component hard sphere fluid \cite{45}. The key idea is high wavevector dominance of the dynamic force correlation vertex in NMCT and the dynamic free energy of NLE theory. The important technical elements are: (1) the wavevector integral below a cutoff $q_c$ can be ignored, (2) for $q \geq q_c$, one can exploit the exact PY theory result \cite{45,49} $c(q) = -4\pi d^3g(d)\cfrac{\cos qd}{(qd)^2}$, and (3) $S_{12}(q)$ is approximated by its high wavevector limit
\begin{eqnarray}
S_{12}(q) = \frac{\rho_1\rho_2 c(q)}{1-\rho c(q)}\approx \rho_1\rho_2c(q).
\label{eq:appendix2}
\end{eqnarray}

Substituting the analytical expressions of $c(q)$ and $S_{12}(q)$ into Eq.(\ref{eq:appendix1}) gives
\begin{eqnarray}
\frac{9}{r_L^2} &\approx& \int_{q_c}^\infty \frac{q^4d{q}}{
2\pi^2}e^{-q^2r_{L}^2/6}\left[\rho_2c^2(q)+\rho_1c^2(q)e^{-q^2r_{L}^2/6} \right] \nonumber\\
&\approx& \frac{24g^2(d)\Phi}{\pi}\int_{q_c}^\infty d{q}e^{-q^2r_{L}^2/6}\left[\alpha+(1-\alpha)e^{-q^2r_{L}^2/6} \right]. \nonumber\\
\label{eq:appendix3}
\end{eqnarray}
One then obtains
\begin{eqnarray}
\frac{\sqrt{3\pi}d}{4r_{L}\Phi g^2(d)}= \left[\sqrt{2}\alpha erfc\left(\frac{q_cr_{L}}{\sqrt{6}}\right)+ (1-\alpha)erfc\left(\frac{q_cr_{L}}{\sqrt{3}}\right) \right].\nonumber\\
\label{eq:appendix4}
\end{eqnarray}
Now, if  $q_cd/\sqrt{3} \ll 1$, the above equation can be explicitly solved:
\begin{eqnarray}
r_{L}(\alpha) = \frac{\sqrt{3\pi}}{4\Phi g^2(d)} \frac{1}{1+\alpha(\sqrt{2}-1) } = \frac{r_{L}(0)}{1+\alpha(\sqrt{2}-1)}.\nonumber\\
\label{eq:appendix5}
\end{eqnarray}

As sketched in section III.C, a similar analysis can be performed for the barrier location and dynamic shear modulus. Straightforward algebra yields for the latter 

\begin{eqnarray}
G(\alpha) &=& \frac{k_BT}{120\pi^2}\int_{0}^{\infty}dq \nonumber\\
&\times& \left(4\pi g(d)d^2 \right)^2\left(\rho_1^2e^{-q^2r_{L}^2/3}+\rho_1\rho_2e^{-q^2r_{L}^2/6} \right) \nonumber\\
&=&\frac{9\Phi k_BT(1-\alpha)}{5\pi d r_L^2(\alpha)}.
\label{eq:appendix6}
\end{eqnarray}

\end{document}